\documentclass[format=acmsmall, review=false, screen=true]{acmart}

\usepackage{booktabs} 
\usepackage{soul}

\usepackage[ruled]{algorithm2e} 

\SetAlFnt{\small}
\SetAlCapFnt{\small}
\SetAlCapNameFnt{\small}
\SetAlCapHSkip{0pt}
\IncMargin{-\parindent}

\acmJournal{TWEB}
\acmVolume{0}
\acmNumber{0}
\acmArticle{0}
\acmYear{2017}
\acmMonth{0}
\copyrightyear{2017}

\setcopyright{acmlicensed}

\acmDOI{0000001.0000001}

\received{February 2017}

\begin{document}
\title[When Simpler Data Does Not Imply Less Information]{When Simpler Data Does Not Imply Less Information: A Study of User Profiling Scenarios with Constrained View of Mobile HTTP(S) Traffic}  
\author{Souneil Park}
\affiliation{%
  \institution{Telefonica Research}
  \streetaddress{Torre Telefónica Diagonal 00 Plaza de Ernest Lluch i Martín, 5}
  \city{Barcelona}
  \postcode{08019}
  \country{Spain}}
\author{Aleksandar Matic}
\affiliation{%
  \institution{Telefonica Alpha}
  \streetaddress{Torre Telefónica Diagonal 00 Plaza de Ernest Lluch i Martín 5}
  \city{Barcelona}
  \postcode{08019}
  \country{Spain}}
\author{Kamini Garg}
\affiliation{%
  \institution{University of Applied Sciences and Arts of Southern Switzerland $\dagger$}
  \department{Informatics}
  \city{Lugano}
  \country{Switzerland}}
 \author{Nuria Oliver}
\affiliation{%
  \institution{Telefonica Research $\ddagger$}
  \streetaddress{}
  \city{London}
  \postcode{}
  \country{United Kingdom}}
  
\begin{abstract}
The exponential growth in smartphone adoption is contributing to the availability of vast amounts of human behavioral data. This data enables the development of increasingly accurate data-driven user models that facilitate the delivery of personalized services which are often free in exchange for the use of its customers' data. Although such usage conventions have raised many privacy concerns, the increasing value of personal data is motivating diverse entities to aggressively collect and exploit the data. In this paper, we unfold profiling scenarios around mobile HTTP(S) traffic, focusing on those that have limited but meaningful segments of the data. The capability of the scenarios to profile personal information is examined with real user data, collected in-the-wild from 61 mobile phone users for a minimum of 30 days. Our study attempts to model heterogeneous user traits and interests, including personality, boredom proneness, demographics, and shopping interests. Based on our modeling results, we discuss various implications to personalization, privacy, and personal data rights.
\end{abstract}

%
%


%
%



\thanks{
$\dagger$  Work carried out as part of her Ph.D. research. Currently in UPC Switzerland.

$\ddagger$ Currently in Vodafone Research.
}

\maketitle

\renewcommand{\shortauthors}{Park et al.}

\section{Introduction}

The exponential growth in smartphone adoption, the availability of thousands of mobile apps connected to the Internet and the development of the Internet-of-Things with billions of connected devices are contributing to the generation of vast amounts of personal data streams. Part of this data is reflective of user activities and behaviors as people carry their mobile devices around-the-clock. The availability of this human-behavioral data combined with sophisticated data-driven machine learning techniques has enabled unprecedented user profiling possibilities. A broad spectrum of players is collecting and analyzing these data streams, such as mobile apps, shops offering free Wi-Fi Internet access to its customers, telecoms and major Internet companies. As useful as it might be for personalizing and improving services, the access to such data streams introduces serious concerns regarding user privacy. 

To date, much progress has been made on understanding the privacy implications of mobile and Internet services that have access to personal identifiable information (PII). However, even without the access to PII, studies have revealed that it is possible to fingerprint \cite{narayanan2008robust}, track \cite{de2013unique}, and carry out different kinds of discrimination \cite{hannak2014measuring} from the analysis of usage logs of specific services, such as browsing history, search logs and movie streaming systems. In consequence, specific legislation and regulation for the treatment of online personal data has been passed in several countries, including Canada and the European Union \cite{ECPPD} aiming to give users back control over their personal data, protect their privacy and simplify the regulatory environment for businesses. This regulation requires Internet services and mobile apps to collect explicit user consent when collecting and analyzing personal data.

Nevertheless, due to the high complexity of the personal data ecosystem and the diversity of entities surrounding the flow of personal data, threats to user privacy (such as data leakage or profiling sensitive information) may arise from entities different than those providing the services with which the users directly interact. For example, an entity positioned in between the user and the service provider might have access to parts of the user's data which --despite being a partial view-- could enable it to profile the user. As a result, users are subject to non-explicit profiling that is beyond the scope of their attention and consent, especially when the observing entities are not intuitively noticeable by the user and/or the data is not clearly subject to privacy policies and personal data laws. The recent change made to the broadband privacy regulation in the U.S. can increase the possibility of such scenarios. This change allows Internet service providers to share private data such as web browsing history without prior user consent \footnote{refer to https://www.nytimes.com/2017/04/03/technology/trump-repeal-online-privacy-protections.html}. 

Our work aims to foster a discussion about non-explicit profiling scenarios by exploring possible profiling approaches with real user data and analyzing their capabilities to profile a variety of potentially sensitive personal information. We refer to the profiling approaches as \textit{constrained profiling} due to the inherent constraints that a profiling entity has. The constraints can stem from multiple reasons, including inherent limitations in the provision of the service (e.g. in the case of HTTPS pages, web browsers can see the content of HTTPS pages while Wi-Fi access points only see up to the domain name of the address), economic limitations (e.g. costs or inability to store large amounts of data), compliance to personal data laws or internal company policies that limit storing certain data for liability reasons. Having awareness of such constraints in practice, we investigate the profiling capabilities of state-of-the-art algorithms when applied to constrained data, and derive implications for designers of user modeling services.

We focus our study on mobile HTTP(S) 
\footnote{Note that we refer to the traffic of HTTP protocol in the following ways: HTTP for unencrypted traffic, HTTPS for encrypted traffic (SSL-over-HTTP), and HTTP(S) for whole HTTP and HTTPS traffic} traffic, and develop possible profiling scenarios considering the potential profiling entities around the data and the constraints they would have. To collect the data, we carried out an in-the-wild user study with 61 participants who gave us access to their mobile HTTP(S) traffic for at least 30 days. The data set includes the traffic generated from any smartphone apps and browsers, which is a relevant data stream for various entities including mobile app providers, telecommunication operators, mobile advertising companies, etc. We apply different constraints to the raw HTTP(S) traffic data - such as having access to only the timestamp of the data, to the header, and ultimately to the full content for the HTTP pages. We consider four profiling scenarios that analyze each of the constrained datasets and also require different levels of technical sophistication, such as the ability to filter noise, categorize websites or analyze Web content. We build user models using state-of-the-art machine learning algorithms taking as input each of the four constraint datasets and compare their performance. In order to cover a wide set of profiling scenarios we consider a variety of personal attributes including personal traits (the Big-5 personality traits and boredom proneness), demographics, and product interests. 

Our study suggests that certain types of personal information inference are still possible even under the constraints described above. The results can be interpreted differently: from the users' perspective, as greater privacy risks from a larger and broader set of profiling entities, encouraging users to be more conservative in sharing data; on the other hand from the perspective of the business providers, as an opportunity to collect less personal data while still being able to provide personalized services. Instead of drawing a one sided interpretation, more fundamentally, our interpretation of the results is to acknowledge the potential value of the users' data and to have 'transparency' as a principle. Although transparency itself does not guarantee a solution\cite{nissenbaum2011contextual, acquisti2005privacy}, we believe that it still is an essential element that enables future discussion about more responsible ways of collecting and using personal data that different parties can agree on.

The main contribution of our study is in unfolding the profiling scenarios around mobile HTTP(S) traffic of people, and examining the capability for inferring various personal information. We discover multiple instances of unexpected profiling scenarios that can be implemented in practice with commonly known techniques, and extract key implications that can shape future discussions. On the other hand, the study is limited in a number of aspects. The data set does not cover the traffic generated in stationary situations, which limits the study having a complete picture of mobile phone usage. In addition, our findings made from the mobile HTTP(S) traffic also suggest more diverse directions for extension considering other diverse behavioral data streams such as location traces and app specific behaviors (\textit{e.g.}, multimedia playlists or photo-taking patterns). The results are also limited in terms of achieving significantly better profiling performance than state-of-the-art methods as our study focuses more on understanding the  user profiling capabilities under a variety of data availability scenarios. 
\section{Related Work}
Many works on personalization and recommendations demonstrate the value of fine-grained online behavioral data for user profiling. Nowadays, many online services (\textit{e.g.} news \cite{liu2010personalized}, music \cite{koenigstein2011yahoo}, and social media feeds \cite{chen2010short}) implement some kind of personalization by using such data. The usage of behavioral data in a wide variety of services reflects how well the data reveals the interests and preferences of people.

Concerns about privacy arise naturally as in some cases the data is informative enough to point out an individual from a pool of users. For example, recent works on de-anonymization found that it is possible to uniquely identify users from movie ratings \cite{narayanan2008robust}, also from mobility traces \cite{de2013unique}. On the other hand, solutions that prevent the identification of individuals have been explored. Research on differential privacy \cite{dwork2008differential} provides means to protect anonymity through generalization or suppression of some attributes of the data \cite{samarati1998protecting} or by adding noise deliberately \cite{iyengar2002transforming}, and at the same time to serve queries up to a certain level of accuracy.

Privacy concerns still remain since the flow of raw behavioral data is complicated, and it is difficult for ordinary users to have awareness or control. A significant portion of Internet services leverage online behavioral data for advertisement as their main monetization strategy. Recent works on analyzing tracking systems in practice reveal a surprising level of exposure of personal online activities to online advertisement platforms \cite{carrascosa2015always}\cite{castelluccia2014selling}. The analyses suggest that a visit to a website is not only visible to the visited website but also to many advertisement platforms. Furthermore, detailed interactions made in the website could be also exposed through cookie-exchange techniques. This information allows service providers to apply sophisticated user-modeling techniques to infer information such as purchase intent \cite{carrascosa2015always}, or price steering \cite{hannak2014measuring}.

The concern intensifies given that such data may reveal or imply other sensitive personal information. Kosinski et al. \cite{kosinski2013private} have analyzed the Facebook Likes of people, and reported that the different Like histories of the people were associated with sensitive personal attributes such as sexual orientation and political views. Lindamood and Kantarcioglu's work \cite{lindamood2009inferring} observed associations between network graph features of Facebook and sensitive personal attributes. Studies have looked at other data as well. Zong et al. \cite{zhong2015you} reported association between the location check-in history and demographic attributes including gender, age, education, and marital status. The kind of apps installed in smartphones was also found to have association with various traits, such as religion, parental status, etc. \cite{seneviratne2014predicting}.

Personality analysis is another area that observed relationships between private information and behavioral data. The study of personality itself is an established area of research, mainly in the psychology literature, and analytical methods based on interviews or questionnaires are already well developed. The 'big-5 personality traits' is a representative example, which measures the personality over the five dimensions,  extroversion, openness, conscientiousness, agreeableness, and neuroticism \cite{goldberg2006international}. Recent studies have found associations between personality and various types of high-level online activities, including overall Internet usage \cite{landers2006investigation}, gaming \cite{jeong2015addictive}, or particular applications such as email, messaging, or social media \cite{tosun2010does}. Ferwerda et al. \cite{ferwerda2016personality} even report the possibility of inferring personality by checking only the disclosed fields of social network accounts. Many works conducted prediction of personality traits with features capturing the technology use, and reported promising performance \cite{de2013predicting, staiano2014money, carrascosa2015always, ramirez2010relationship, tsao2013big}. 

Our work expands the scope of the literature by considering ancillary entities that are unrelated to users' interaction but still have certain level of access to the behavioral data (\textit{e.g.}, mobile apps, WiFi access points, ISP and telecommunication operators, mobile advertising companies). The capacity to profile users is likely to be limited inherently due to the different restrictions imposed on the entities, however, it is unclear what types of and how much profiling  can be performed by them. As the current work is specific to HTTP(S) traffic from mobile phones, we believe there can be additional future works with similar goals as ours but which explore different types of behavioral data.

\section{Data and Constraints}
In this section we describe the scope of our data and the four different scenarios that we have analyzed in our study as a result of applying four constraints to the raw HTTP(S) data.

The data collection was conducted through an web acceleration proxy infrastructure for mobile devices that all the participants in our study configured in their smartphones. Their HTTP(S) traffic went through the proxy when they were connected to the Internet via cellular networks (2G/3G/4G). It is important to mention that the proxy does not archive the content of the HTTP(S) traffic. Instead it keeps the URL, and in the case of HTTPS, only the domain name. Thus, the content analyses conducted in the later part of the study use a limited portion of the pages that could be retrieved solely with the URL, i.e., publicly accessible pages. The private pages such as the HTTPS content, those accessed with credentials or tokens, or those that are personalized through cookies are not retrieved and excluded in the content analyses.

We apply different constraints to the raw HTTP(S) traffic which are reflective of having access to (a) explicit user consent or not; (b) different levels of visibility into the data and (c) different capabilities in technological sophistication. In this regard, we devise four scenarios to infer user traits, attributes and interests from this constrained data. 

As mentioned, there are various entities which can have access to the HTTP(S) traffic and potentially perform profiling. While there are many studies \cite{enck2011study} that deal with mobile apps particularly, and use the data that only they can access (e.g., sensor data, GPS readings, and personal data including contacts and photos), our work takes a different perspective. We deal with a particular data set that is of interest to many entities, i.e., HTTP(S) traffic, but explore various types of profiling entities.

In our study, we focus on analyzing the profiling capability of each scenario separately rather than speculating on possible combinations of different scenarios. It is possible to think of a combination of different scenarios since there are cases where an entity with less constraints can perform an analysis of another entity which has more constraints. For example, a web browser can not only analyze the patterns in the visited URLs but also see the page contents. However, such combinations may not be generalised as the profiling entities are under different circumstances. For example, while the web browser can see the page contents it cannot see the HTTP(S) traffic of other applications where ISPs or proxy servers can see those traffic to some extent. As the combinations can take different forms depending on different situations, we believe that it is essential to first understand each scenario more deeply since the results can also serve as a reference for possible combinations.

\subsection{Profiling Scenarios around HTTP(S) traffic}
\begin{figure}
  \includegraphics[scale=0.6]{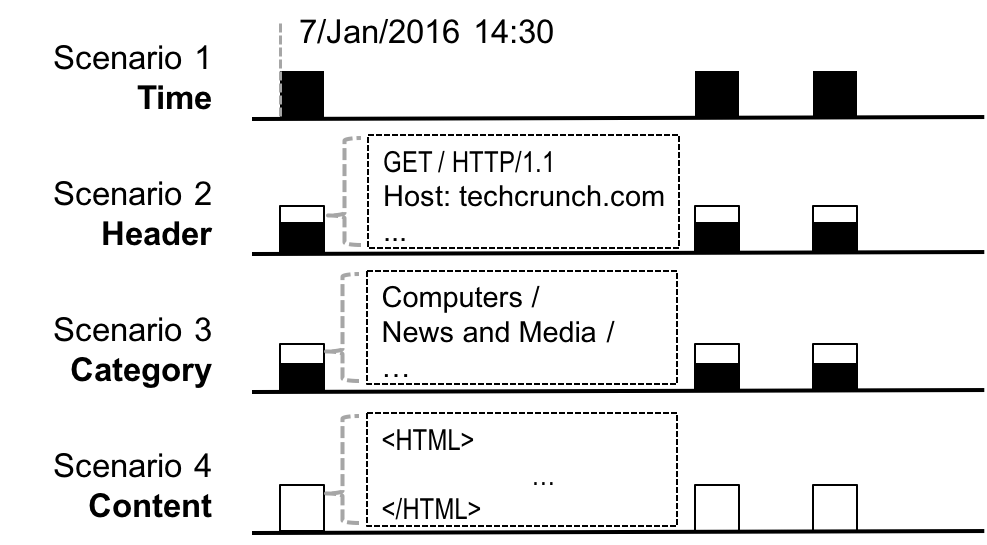}
  \caption{Four user modeling scenarios by applying four constraints to HTTP(S) mobile traffic.}
  \label{fig:one}
\end{figure}

Next we describe the four different profiling scenarios, depicted in Figure~\ref{fig:one}.

\paragraph{Scenario 1 (S1) - Profiling based on time-stamps} 
The strictest constraint we assume is such that the S1 data is composed only of the timestamp of the HTTP(S) accesses. It assumes no availability to the content of the HTTP(S) message or processing of the HTTP(S) header. Even though the resulting data seems very simple, the stream of timestamps may carry meaningful behavioral information about the user's habits and routines. For instance, in our collection scenario, S1 data might be reflective of time-periods when the user is outdoors or on-the-move as we only collect and analyze data going through cellular networks\footnote{Note that this is an illustrative example. We do not fully rely on this assumption in our analysis as mobile Internet can be also used indoors (or Wi-Fi outdoors)}. 
Note that all mobile apps on Android could acquire S1 data by tracking when a user is connected to the mobile Internet or to Wi-Fi \footnote{refer to https://developer.android.com/training/basics/network-ops/managing.html}.
(typically in order to prompt updates only on Wi-Fi). We believe that profiling methods that analyze S1 data can be easily built in practice as capturing and storing the timestamps of HTTP(S) traffic is becoming a feasible process for a myriad of players in the Internet ecosystem (for instance, end-devices with limited computing power and storage capacity --such as IoT devices). 

\paragraph{Scenario 2 (S2) - Profiling based on header} 
The second scenario refers to the cases in which only the \textit{header} of the HTTP(S) traffic is available. Whereas the previous constraint only reveals the existence of HTTP(S) traffic, the \textit{type} of user activity becomes accessible under this scenario. The header reveals important information related to the user's activity including the destination address, the amount of data exchanged, and the app used for access in some particular cases (through the user agent field). It evokes profiling scenarios that exploit patterns in the usage of apps or in accessed domain names. The entities that participate in the delivery of HTTP(S) traffic between the user and the destination - such as ISPs, mobile telecommunication operators, and web proxy servers - could analyze this kind of data. Depending on the  platform and versions, there are mobile apps which also have access to the traffic information\footnote{The apps for network usage statistics use such data. An example is available at http://network-connections.mobi/}. 

\paragraph{Scenario 3 (S3) - Profiling based on domain name} 
The third scenario assumes the ability interpret the topical categories of URLs, for example, 'Computers/News and Media' for the domain name 'techcruch.com' Although this may not seem to add much data to the HTTP(S) headers, the difference is significant as it requires merging the HTTP(S) logs with external sources to identify the topical category of the URLs. The scenario is also a critical stage where semantic interpretations of the logs and profiling of preferences become possible. 

Today, there are several online tools and resources that support the categorization of domain names, which makes the implementation of the scenario practical. These tools include open website dictionaries (e.g., Dmoz.org) that are built in a crowd-sourced manner, and machine learning tools that assume a certain topical category and perform a classification of websites with some example pages. As it is impossible to recognize and categorize all individual pages available on the Web, the categorizations are often made at an aggregated-level, such as at the domain name-level. Though the categorizations do not exactly capture the actual delivered content, such tools allow a profiling method to guess an approximate topical category based on popular websites. In fact, this approach is often used in computational advertisement, where the user profiles built from website categories are used to determine the displayed ads or price offerings for each user \cite{castelluccia2014selling}. We therefore make this scenario as a separate one from the aforementioned scenarios. Note that the data in this scenario can also deal with HTTPS traffic given that the domain name is typically visible regardless of the content encryption. 

There are products that demonstrate the types of entities that have access to such information. Prior works on real-time bidding \cite{carrascosa2015always} shows that ad exchange platforms (e.g., doubleclick) perform personalized advertisements based on web browsing patterns. Similar products have been also offered by telecom operators\footnote{Refer to http://www.multichannel.com/news/advanced-advertising/att-drop-internet-preferences-program-gigapower/408181 and https://www.verizonwireless.com/support/verizon-selects/}. 
 In Android, there have been APIs for querying the browsing history of the device\footnote{Though removed from Android 6.0, the permission com.android.browser.permission.READ\_HISTORY\_BOOKMARKS allowed fetching the browser history.}, which allow mobile apps to have access to similar data.

\paragraph{Scenario 4 (S4) - Profiling based on page content} 
The final constraint on the HTTP(S) traffic assumes access to the actual HTTP pages as the full URL is exposed. This type of data enables profiling methods that incorporate advanced content analysis techniques. A profiling entity can implement such a method either by archiving the page content of the HTTP accesses or by storing the full URL and then fetching the documents later for analysis and profiling purposes. 

While this constraint provides the maximum level of detail regarding the \textit{content} that the user has accessed, it is not available for HTTPS traffic where the URL is encrypted and only the domain name is available. Note that collecting and analyzing this type of data requires large-scale storage capabilities. Moreover, explicit user consent may be necessary and strict personal data protection laws might apply to this type of data given the potential sensitivity of the accessed content. In theory, the entities that are mentioned for scenario 3 have the capability to implement this profiling scenario. 

The constraints defined above are specific to mobile HTTP(S) traffic, which is the subject of our study. One may define different constraints for other types of data, user modeling goals and application environments. We leave to future work the exploration of the concept of constrained profiling in other domains beyond mobile HTTP(S) traffic.

\section{Study Design}
We conducted an in-the-wild user study to collect: (1) mobile HTTP(S) traffic, and (2) self-reported traits and interests about our participants through a series of online questionnaires. The HTTP(S) traffic was collected over 6 months during which we logged an average of 77 $\pm$ 21 days per participant. We describe the participants and the data collection methodology below.

\subsection{Participants}
We recruited ~200 smartphone users in Spain. They were recruited through two channels: first, an email list of volunteers who signed up for participation in user experiments carried out by our research organization. Second, a recruitment agency for user studies. The participants did not receive fixed incentives, but we held monthly raffles during the data collection period (from August 2015 to January 2016) for a 100 Euro gift card of a large online store. Participants were enrolled via the study's website which guided them to go through three stages: (1) a consent form
\footnote{Due to the difference in the legal and organizational infrastructure of the country where the study took place, the study did not go through the Institutional Review Board review process. The Ethics Committee, which is a similar institution in the European Union, deals with clinical trials.}
, (2) several user profile questionnaires to collect ground truth, and (3) instructions on how to configure their mobile phones to use our mobile proxy. 

As the study collects potentially sensitive information, we elaborated on the data collection and our analyses throughout the consent form in a reader-friendly way while assuming a non-technical audience. The consent form described upfront that the study aims to explore the relationship between mobile browsing behaviors and personal characteristics. It further described that we would explore various data mining techniques applied on the collected data and that no human will read their logs. We also made a dedicated section to inform about the data management process. It described that the browsing history would be collected from their mobile device, and we would not collect any other personally identifiable information than the email address. We also clarified that the data would not be shared outside the research organization nor leave, and that the data would be deleted once the study is completed. The data was stored in a way that allowed access to the researchers involved in this study only via the intranet of the organization.

\begin{figure}
  \includegraphics[scale=0.6]{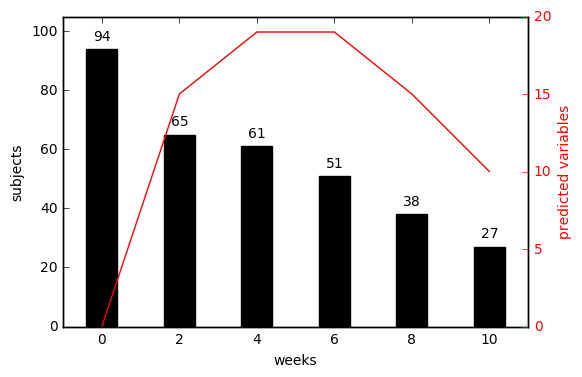}
  \caption{Trade-off between the subjects and the data availability: the bars show the number of subjects who satisfy the data availability requirement (x-axis), and the red line shows the number of target variables predicted with a certain level of balanced accuracy ($>$ 60\%).}
  \label{fig:zero}
\end{figure}

Only the applicants who successfully completed the three stages were enrolled to the study, namely the ones who consented, filled out the questionnaires and properly configured our mobile proxy. While 94 out of the 200 participants successfully completed these three stages, the amount of collected data per participant varied from 8 days to 170 days. Hence, we had a trade-off between maximizing the number of days of the collected logs per user and the number of available subjects who satisfied the threshold. We iteratively tested a range of the thresholds and observed that the accuracy of our models stabilizes with at least 30 days (61 subjects) of internet logs. Interestingly, the modeling performance remained similar as we were increasing the threshold of the number of days, however, sharply dropped when less than 40 subjects were involved in the analysis. Lowering the threshold below 30 days also negatively affected the performance as it adds noise, potentially suggesting that less than one month of logs does not suffice to extract features that quantify users' typical internet usage routines. Figure~\ref{fig:zero} shows the details about the trade-off. We thus present the results taken from the 61 participants who contributed their data for at least 30 days. Table~\ref{tbl:two} summarizes our participants' demographic information.

\begin{table}
  \includegraphics[scale=0.5]{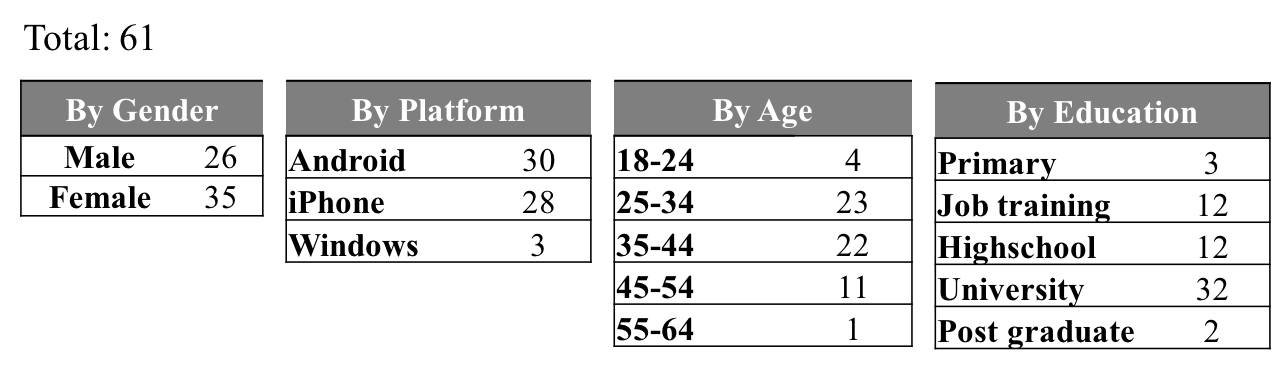}
  \caption{Demographics of the participants in our study.}
  \label{tbl:two}
\end{table}

\subsection{Mobile Online Activity Data Collection}
The proxy-based approach that we followed for data collection has several advantages. First, it does not require participants to install any app and spend resources of their mobile device for the study. Second, we collect the HTTP(S) traces regardless of the app or browser that generated them. As many apps use the HTTP(S) protocol to communicate, our approach can capture information about the usage of apps. In addition, we log web page accesses from any app, for example, a page opened with a non-default browser by clicking a link in the Twitter app. Finally, it does not require to write and maintain several versions of the data collection software for different mobile operating systems. Note that our approach is limited when it comes to capturing the activities inside mobile apps, such as tweeting inside the Twitter app. This limitation is difficult to address since apps usually have their own internal protocol and data format. With respect to HTTPS traffic, we do not obtain the full URL of the destination but only the hostname. In addition, recall that we do not collect private pages that require tokens or authentication. 

\subsection{Questionnaires}
We considered various aspects that could be of interest of service providers for personalizing individual online experiences. In this regard, the participants filled multiple questionnaires to collect ground truth information about their characteristics and interests, which we aimed to model. We asked the participants to report their shopping interests (being an evident focus for online advertisers), demographics (that represents one of the basic categories of variables when it comes to using computers and internet), and finally personal traits (that have been shown to be relevant to UX personalization). Specifically, we collected the following information: 
\paragraph{\textbf{1) Demographics}} Participants provided us with their gender, age and education level as depicted in Table~\ref{tbl:two}.

\paragraph{\textbf{2) Personality}} We collected our participants' Big-5 personality traits using the widely validated 50-item IPIP questionnaire \cite{goldberg2006international}. This model is commonly used in Social Psychology to characterize personality using five dimensions (extraversion, neuroticism, agreeableness, conscientiousness and openness). Over the past decade the Big-5 has attracted the attention of the user modeling community due to the impact of an individuals' personality on their mobile phone usage \cite{butt2008personality}, online activities \cite{ramirez2010relationship}, satisfaction with technological services \cite{oliveira2013influence}, and online purchasing preferences \cite{huang2010relationship}.

\paragraph{\textbf{3) Boredom Proneness}} We quantified the tendency of individuals to experience boredom using the Boredom Proneness Scale \cite{matic2015boredom}, which has been used for this purpose for the last three decades. Boredom proneness is associated with susceptibleness to stimulation seeking behavior and mobile phones are often used as a stimulation source \cite{barabasi2005origin}.

\paragraph{\textbf{4) Shopping Interests}} We obtained self-reported information about product interests by asking our participants about basic purchase frequency of various products through questions on a 7-point Likert scale (1=never, 7=very often). We identified 14 product categories commonly listed in online stores: books, computers, software, mobile apps, music, videos, flowers, flights, tickets, clothes, travel, furniture, home appliances, and groceries.

\section{Feature Extraction}
We compute a rich set of features for each data type with the goal of capturing the users' mobile online activity routines. We elaborate on the features below, whose summary is shown in Table~\ref{tbl:one}. 

\begin{table}
  \includegraphics[scale=0.5]{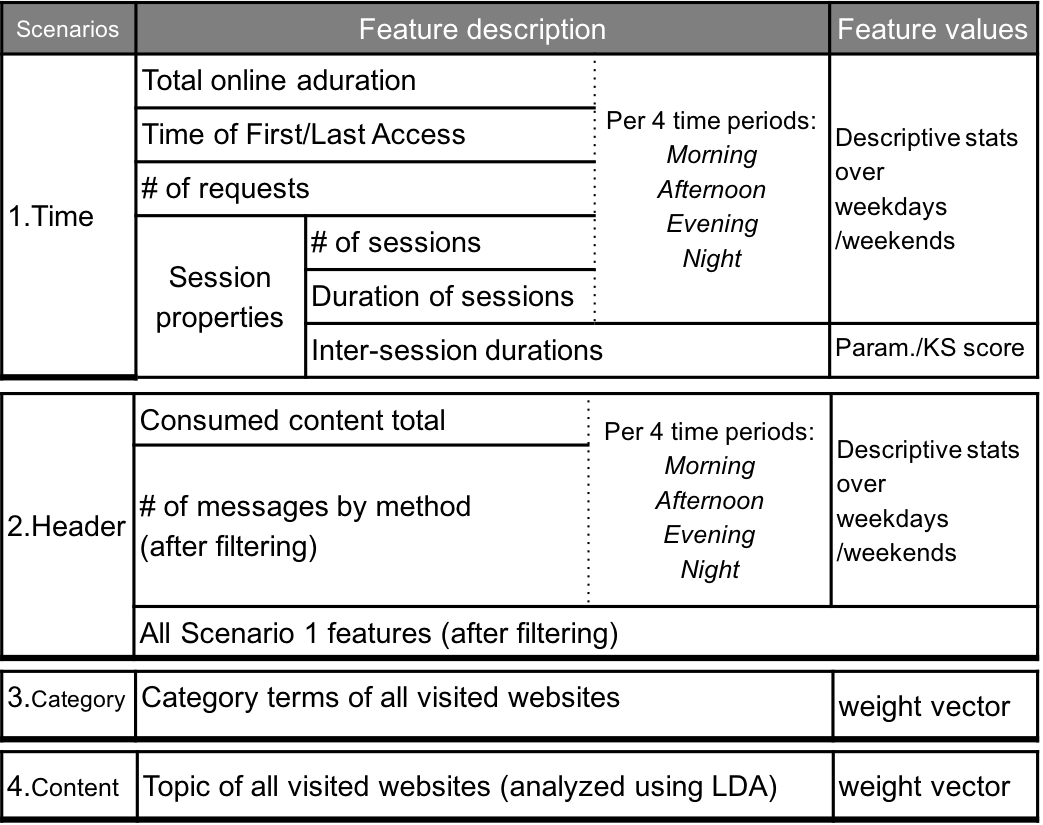}
  \caption{Overview of the developed features.}
  \label{tbl:one}
\end{table}

\subsection{Scenario 1 (S1): Dynamics of HTTP(S) Access}
The timestamp information of HTTP(S) traffic can be obtained without looking into the HTTP(S) message at all. The features at this level are designed to capture the temporal patterns in mobile online accesses; patterns derived from when an online access starts, how long it proceeds for, and how often such accesses take place. Note that S1 data includes not only user-initiated HTTP(S) requests but also automatically generated requests regardless of actual user interaction (\textit{e.g.}, background activity of apps or OS platform-level communication). As we do not look into any field of the HTTP(S) traffic in this scenario, we are not able to distinguish and filter them out. 

To compute features from the data of scenarios 1 and 2 (S2, explained below), we divide the day into \textit{four time periods} and compute features aggregated per time period: morning (5am to 12pm), afternoon (12pm to 6pm), evening (6pm to 10pm) and night (10pm to 5am).
We compute the following features:

\paragraph{1. First and last access time} We take the time of the first and the last HTTP(S) requests for each period of the day. 

\paragraph{2. Total number of requests} This feature counts the number of observed requests for each period of the day.

\paragraph{3. Total online duration} We approximate the total duration of mobile online activity per day and per period of the day. For this, we break down the day into 10 minute slots, and then count the slots when there is an HTTP(S) traffic. 

\paragraph{4. Estimation of sessions, their frequency and duration} Using a time window of 5 minutes, we estimate sessions by grouping the consecutive requests that are made within the time window. We then count the number of sessions and measure the total duration of the sessions, again for each of period of the day. 

\paragraph{5. Inter-session time} We measure the time between every two consecutive sessions of a day. We collect all the measurements over the whole study period, fit to Weibull and Power-Law distributions
\footnote{In addition to the parameters of the two distributions, we included also Kolmogorov-Smirnov scores as one of the features.}
 and compute the distribution parameters. This is in accordance with a number of studies to characterize burstiness in a variety of human behaviors \cite{barabasi2005origin} and particularly phone usage patterns follow one of the two distributions \cite{jiang2013calling}.

Once the features are extracted per each day, we compute descriptive statistics separately for weekends and weekdays
\footnote{More precisely working vs non-working days (the latter including weekends and bank-holidays).}
. Overall, we compute 135 features in this scenario. 

\subsection{Scenario 2 (S2): Header Analysis}
In this scenario, we assume that the data is constrained only to the header of the HTTP(S) messages - not the content. Among the fields in the header, we take into account the three fields that could be related to the activity of the users: request method (i.e., HTTP(S) verb), delivered data, and host and path which informs the destination URL. These fields inform where the users visit, the amount of content consumed (bytes), and possibly the type of interaction such as POST, GET, or CONNECT for HTTPS. Except the full path of the URLs, the fields can be observed also for HTTPS traffic
\footnote{As for HTTPS traffic, the field 'delivered data' is computed by taking the overall sum of bytes transferred via a TLS session.}
. The other fields mostly depend on the underlying infrastructure rather than on user actions. 

Another advantage over the first scenario is that it opens the possibility to identify and filter out automatically generated traffic. As mentioned, the automatic traffic could be generated by the background activity of apps, operating system, and also by web browsers to fetch objects for page rendering. Filtering such traffic helps the analyses to reflect the actual behavior of users more accurately and avoid possible biases due to certain apps or platforms. While the filtering task itself is a challenging research problem, we implemented a simple and practical solution using an HTTP(S) client. It filters out the HTTP(S) requests that do not return an web document with certain content (text, image, or video), and those made to page objects and resources (e.g., json, css, and Javascript files) since they are likely to be triggered by browsers automatically in the rendering process rather than by a user's request. 

We compute the following features from data in S2: 

\paragraph{1. Total amount of consumed content} This feature sums up the total number of bytes exchanged during each period of the day. This feature is calculated using all the messages regardless of whether they are generated through user interaction or not since we are interested in the total amount of generated traffic. 

\paragraph{2. Number of messages by method} For each HTTP(S) verb, we count the number of messages exchanged in each period of the day. In this case, we only consider the messages that are associated to user interaction.

\paragraph{3. Features in S1 after filtering} We compute all the features from S1 data but only with the messages that are associated with user interaction.  

Similar to the data in S1, we compute descriptive statistics separately for weekends and weekdays yielding a total of 180 (135 + 45) features.

\subsection{Scenario 3 (S3): Category of web pages}
As described, this scenario requires a method for identifying the topical category of a given URL. We use DMOZ
\footnote{www.dmoz.org}
, a commonly used open directory of websites, to annotate the destination hostnames with semantic tags. First, for each participant, we sort out the HTTP(S) messages that are associated with user interaction as we do for Scenario 2. Second, we query the DMOZ directory for all the messages in order to obtain their topical categories. The dictionary returns a category of multiple hierarchies, for example,  'Computers/News and Media' for the website 'techcruch.com' We take into account the categories of all the hierarchies for feature construction. 

We generate the features by creating a term vector for each participant, with as many dimensions as categories and a weight value assigned to each category. We tried two weighting schemes: term frequency (TF), which counts the frequency of categories from the history of hostnames for each participant; and TF-IDF, which counts the frequency of the category from the history of hostnames (the TF term), and then scales the frequency based on the commonality of the corresponding category across all participants (the IDF term). It emphasizes the categories that distinguish a user from the others. The IDF term decreases the weight of the categories that commonly appear across subjects. 

We additionally apply Latent Semantic Indexing (LSI) to reduce the dimensionality of the vectors due to the sparsity of the original term vectors. LSI is a popularly used technique especially in information retrieval, which reduces the dimensionality by analyzing the similarity between the original dimensions using singular vector decomposition. 

\subsection{Scenario 4 (S4): Topic of web pages}
The data analyzed in S4 assumes having access to the content of a limited set of web pages that are publicly accessible. 

We create a topic profile for each participant from the content of the visited web pages. In order to recognize the topic of the web pages, a topic model is built first using the content of all the web pages accessed by all the participants. The topic model identifies the major topics observed in the corpus, which serves as a categorization framework of web pages. Once the major topics are identified, we use the topic model to estimate the topic distribution of every web page over the identified major topics, and compute the topic profile of each participant by aggregating the topic distribution of the browsed pages. 

The corpus of HTTP pages goes through three pre-processing steps before the topic model construction: main text extraction, language recognition, and lemmatization. In order to filter noise and focus on the main content, we use Boilerpipe \cite{kohlschutter2010boilerplate}, a web page parsing library. As our corpus includes web pages of different languages, language identification is crucial for the subsequent analysis process. The language of each web page is identified through Libtextcat \cite{cavnar1994n}, and the pages of the top two most common languages (Spanish and English) are taken for analysis as they cover the majority of the corpus. Finally, we use FreeLing \cite{carreras2004freeling} for lemmatization of both Spanish and English text and perform stop-word removal. A topic model is created separately for the two languages. 

We use Latent Dirichlet Allocation (LDA) \cite{blei2003latent} to build the topic model. LDA is a fully Bayesian unsupervised framework for inferring latent topics of a given corpus. It views documents as mixtures of topics and topics as mixtures of words. During topic inference, both sets of mixtures (words and topics) are adjusted to maximize the likelihood of the input corpus. 

We implement LDA based on Collapsed Gibbs Sampling. The two hyper-parameters $\alpha$ and $\beta$ are tuned using the Digamma Recurrence Relation \cite{breiman2001random}. The number of topics, \textit{K}, is set to 20 (per language), as it provides a good trade-off between topic specificity and coverage in our application setting. After applying LDA, we manually label each topic from the keywords produced by LDA and by looking at sample web pages associated with the topic.

\section{Evaluation Setup}
Next, we build machine learning-based binary classifiers for each of the target variables (i.e. personality, boredom proneness, demographics, and shopping interests) and for each of the data types from scenarios 1 through 4. In order to be able to carry out binary classification, we split participants into two sets (low vs. high) using the median value of each target variable. This approach has been used frequently in the literature \cite{chittaranjan2013mining, de2013predicting, staiano2014money}. As multiple participants can have the median value and the median based split does not evenly divide the participant pool accordingly, the baseline accuracies are sometimes not exactly 50\%. 

We perform a quantitative comparison of the quality of the user models built in each scenario and derive insights about the relationship between mobile online activity and user traits/interests. Since several related works have used a similar approach \cite{chittaranjan2013mining, de2013predicting, staiano2014money}, we can also interpret the quality of our results by comparing them to those reported in the literature. 

As mentioned earlier, the features computed in each scenario are evaluated separately. Recall that having access to the hostname and/or the content of the HTTP(S) message (S3 and S4) does not necessarily imply being able to access the data considered in S1 and S2. For example, a web browser client can access data from S3 and S4 but typically lacks the ability to access the overall mobile-phone traffic (such as including app-generated traffic) in order to compute features corresponding to S1 and S2.

\subsection{Ground-truth}
Next, we report the statistics of the ground truth variables obtained from our participants' answers to the questionnaires.

\subsubsection{Demographics}
Table~\ref{tbl:two} shows the demographic distribution of the participants. As seen in the Table, our participants are diverse in gender, age and education levels, which supports a general interpretation of our results.

\subsubsection{Personality}
The answers to the IPIP Big-5 Personality Test showed high internal consistency, with Cronbach's alpha values of .78 (openness), .81 (extroversion), .80 (conscientiousness), .87 (agreeableness), and .86 (neuroticism). The scores were also near-normally distributed. Though we were not able to verify if the statistics (mean and st. dev) of our sample (shown in Table~\ref{tbl:three}) match with those of a larger Spanish population due to the lack of relevant literature, we believe that our results are reliable as the internal consistency matches that of prior Big-5 tests. The mean internal consistency of the IPIP Big-5 test \cite{goldberg2006international} is .84. In addition, Del Barrio et al. \cite{del2003personalidad} also reported the results of a Spanish language version, with Cronbach's alpha between .78 and .88 and temporal stability (r between .71 and .84). 
The median value-based division for the binary classification task gave two balanced sets for all the five personality dimensions: 31 participants (\textit{high}) vs. 30 participants (\textit{low}) for Extraversion, Agreeableness, and Neuroticism, and 33 participants (\textit{high}) vs. 28 participants (\textit{low}) for Conscientiousness and Openness.  

\begin{table}
  \includegraphics[scale=0.5]{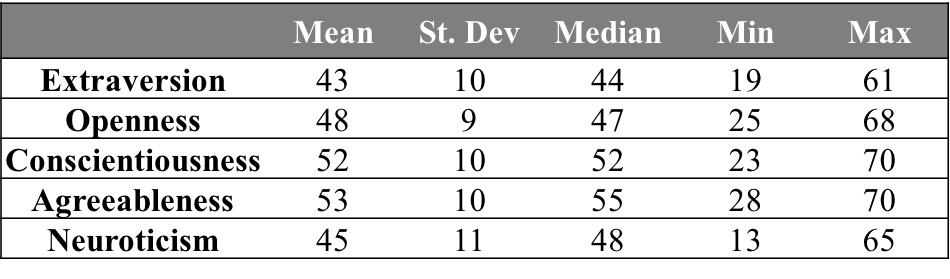}
  \caption{Big-5 statistics of our participants}
  \label{tbl:three}
\end{table}

\subsubsection{Boredom Proneness}
The answers to the Boredom Proneness Scale also demonstrated a high internal consistency with an alpha value of .86 --which is in accordance to previous work \cite{farmer1986boredom} and demonstrates satisfactory levels of internal consistency (alpha = .79) as well as of test-retest reliability (r = .83). The median value-based division for the binary classification task also gave two balanced sets with 31 and 30 participants in the high and low boredom proneness classes respectively. 

\subsubsection{	Shopping Variables}
While we inquired the frequency of purchasing for 14 product categories, we excluded five of them from the evaluation since the responses were extremely skewed to the selection "never purchase it". For the other 9 categories, we segment participants into two sets using the median value per category. Table~\ref{tbl:four} depicts the statistics for the four example product categories for which the classification accuracy in our experiments was sufficiently high.

\subsection{Model Selection}
We tested a number of machine learning-based classifiers and chose Gradient Boosting Machines (GBMs) in all reported results. Three different classes of algorithms were tested: first, Support Vector Machines (SVMs) (with RBF and linear kernels) given their high performance in related tasks \cite{chittaranjan2013mining, matic2015boredom, de2013predicting}; second, decision tree-based methods (namely, Random Forests and GBMs) since they satisfy the max-margin property, yield state-of-the-art results and do not require feature space specification \cite{breiman2001random}; third, probabilistic methods (Na{\" i}ve Bayes). 

\begin{table}
  \includegraphics[scale=0.5]{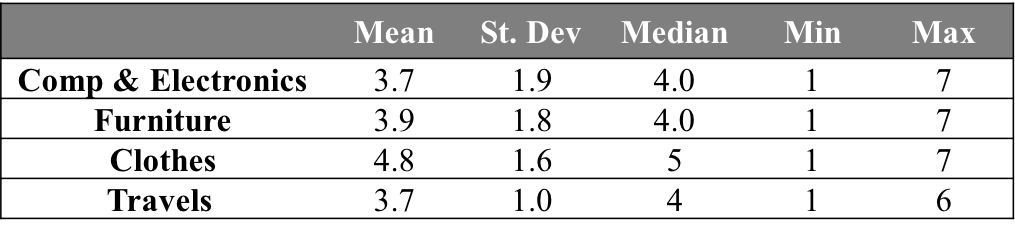}
  \caption{Selected Shopping Interest - Statistics}
  \label{tbl:four}
\end{table}

This testing process helped us approach the task from different angles. GBMs and SVMs with RBFs kernel outperformed others, and occasionally SVMs were more accurate than GBMs. However, we chose GBMs as they showed more stable performance across the target variables and they do not require feature space specification. Thus, they are not affected in their performance by feature selection. 

Our implementation of GBMs is based on the R library \textit{XGBoost}
\footnote{https://xgboost.readthedocs.org/en/latest/}. 
We tried a set of parameter combinations to prevent overfitting: \textit{eta} that determines the learning rate, \textit{gamma} regulating the sensitiveness to training examples, and the number of iterations. The set of combinations we tried includes the default values defined by the library, 0.3 for \textit{eta} and 0 for \textit{gamma}, and the ones with lower \textit{eta} (0.1 or 0.2) and higher \textit{gamma} (2 or 3) which can help avoiding overfitting. For the combinations with lower \textit{eta} we doubled the number of iterations as the learning rate is slower. For each data type, we applied the same set of combinations and chose the one with the best performance across all target variables. Considering the number of instances (61) we measured performance through a leave-one-out approach, which sequentially selects one data point (i.e. in our case one participant), trains the model with the rest of the data points and tests the model with the selected data point.

\section{Results and Discussion}
We first provide an overview of the results followed by a detailed description of the most predictive features for each target variable. 

\subsection{Results Overview}

\begin{table}
  \includegraphics[scale=0.45]{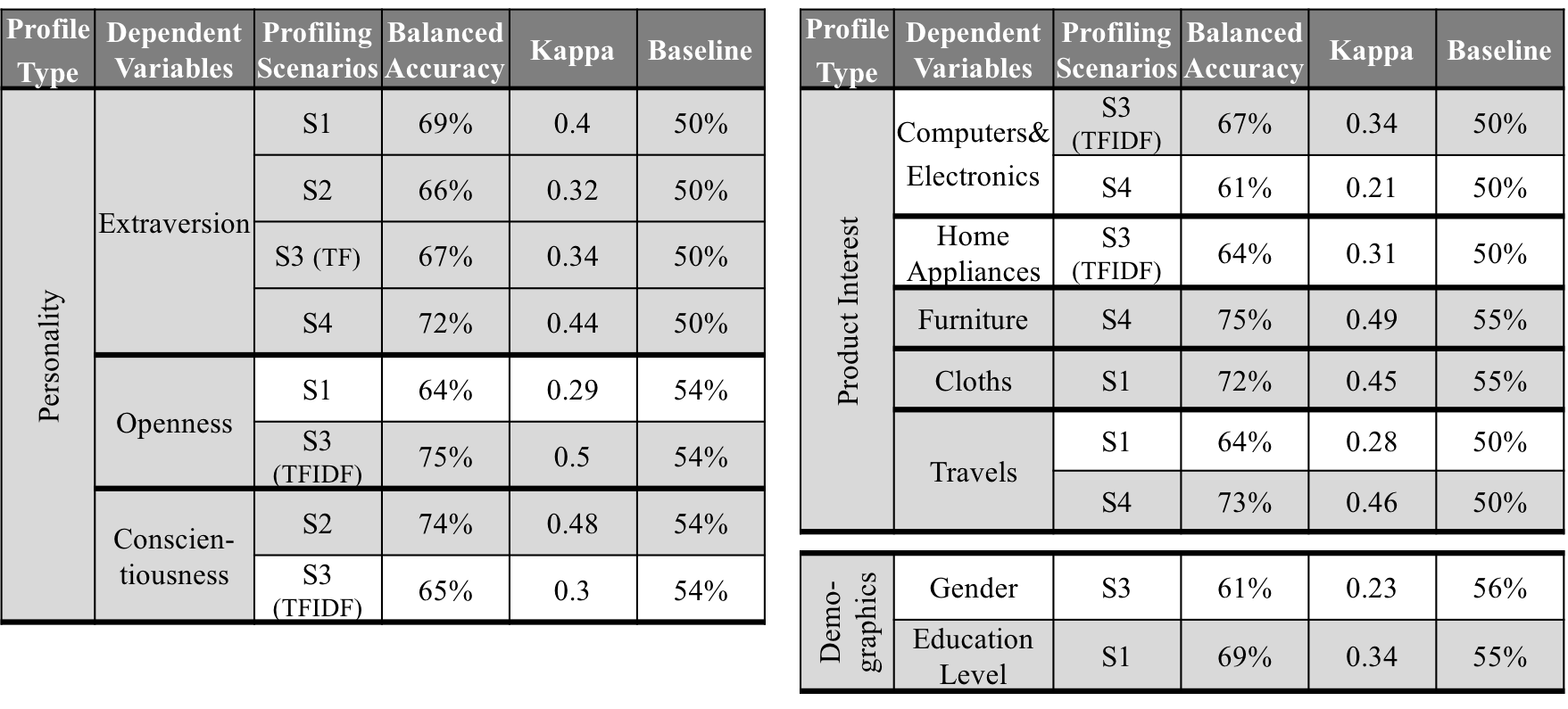}
  \caption{Classification Accuracies of all Models (balanced accuracy between 60\% and 65\% is in white, above 65\% is in grey)}
  \label{tbl:five}
\end{table}

Table~\ref{tbl:five} provides an overview of the results. As the analysis includes a large number of target variables, we only report the cases with a performance competitive with respect to the state-of-the-art performance reported in previous works for similar tasks using mobile data. We report results with a balanced accuracy \footnote{balanced accuracy is a metric for evaluation of classifiers which deals with class imbalance. It is the arithmetic mean of sensitivity ($\frac{\text{true positives}}{\text{true positives} + \text{false negatives}}$)and specificity ($\frac{\text{true negatives}}{\text{true negatives} + \text{false positives}}$).}
between 60 and 65\% (Table~\ref{tbl:five} in white), and with a balanced accuracy
\footnote{The baseline random model refers to a random binary guess accuracy taking into account unbalances in two classes. In addition, to provide a better picture of the classification accuracy we provide also the related confusion matrices (Table~\ref{tbl:five}).
} over 65\% (Table~\ref{tbl:five} in grey). Note that all results have a kappa value between 0.21 and 0.5, meaning that the learned models are 21\% to 50\% better than a baseline random model. While our results do not outperform those reported in the literature, note that our research focuses on examining the profiling capability of the possible scenarios in practice rather than on developing a more accurate method. 

\paragraph{Scenarios 1 and 2} Perhaps the most surprising finding has been the modelling power of the features computed from S1 and S2 data, and particularly from S1 data given that it characterizes overall temporal patterns of online traffic, independently of whether it is user-initiated or not. Even though this might seem to be a very coarse and noisy signal, the results show that it is possible to infer user traits of different nature: personality (extraversion with 69\% balanced accuracy), demographics (educational level with 69\% balanced accuracy) and even a purchasing interest in clothes (with 72\% balanced accuracy) and traveling (with 64\% balanced accuracy). As the non-user initiated traffic is filtered by applying constraint 2, another personality variable, conscientiousness (with 75\% balanced accuracy), is inferred with features from S2 data. 

\paragraph{Scenarios 3 and 4} As expected, having access to the content's semantic information enables us to infer personal preferences. Although this finding may seem obvious, it is important to note that in this work we only analyze the HTTP(S) traffic through the 2G/3G/4G mobile network, which is a subset of the entire online activity of an individual. Despite this limitation, it is possible to infer our participants' level of interest in several product categories: computer \& electronics (67\% balanced accuracy), furniture (75\% balanced accuracy), travel (73\% balanced accuracy), and home appliances at a lower (but better than not-random) balanced accuracy (63\%).

Interestingly, the models built with features from S3 and S4 data did not classify personality traits and demographics better than the models built with features from S1 and S2 data. However, we are careful in reading this finding since there might be other reasons for this performance that are not related to the nature of the data, such as missing entries in the website dictionary (S3 data) or the limitation of the scope of the pages that we could retrieve through the simple HTTP client (S4 data).

While it is difficult to compare the performance for all the target variables to those of the literature, we compare the results for the personality traits and boredom proneness. We additionally present the comparison in Table~\ref{tbl:six}. From the call logs, de Montjoye et al. \cite{de2013predicting} detected all the five personality traits with 49\% to 63\% accuracy (with the baseline being 36\% to 39\%). Smartphone usage and sensor logs were used in \cite{chittaranjan2013mining} to detect the five personality traits with the F-measure ranging between 0.6 and 0.8. Staiano et al \cite{staiano2012friends} explored social network modelling (using both call logs and mobile phone sensors) and detected the five personality traits with 62\% up to 71\% of accuracy (with the baseline accuracy being around 50\%). There is only one study that modelled boredom proneness \cite{matic2015boredom} and achieved the accuracy of 80\% in a binary classification.

\begin{table}
  \includegraphics[scale=0.6]{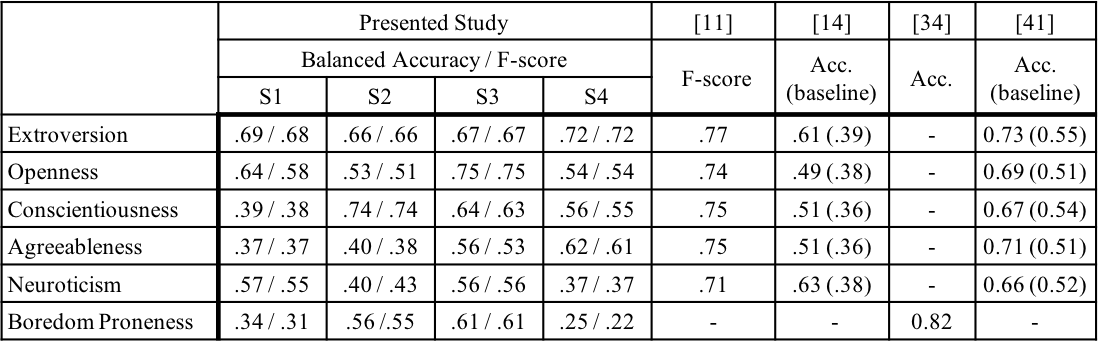}
  \caption{Comparison to related studies}
  \label{tbl:six}
\end{table}

\subsection{Most Predictive Features per Model}
\begin{table}
  \includegraphics[scale=0.5]{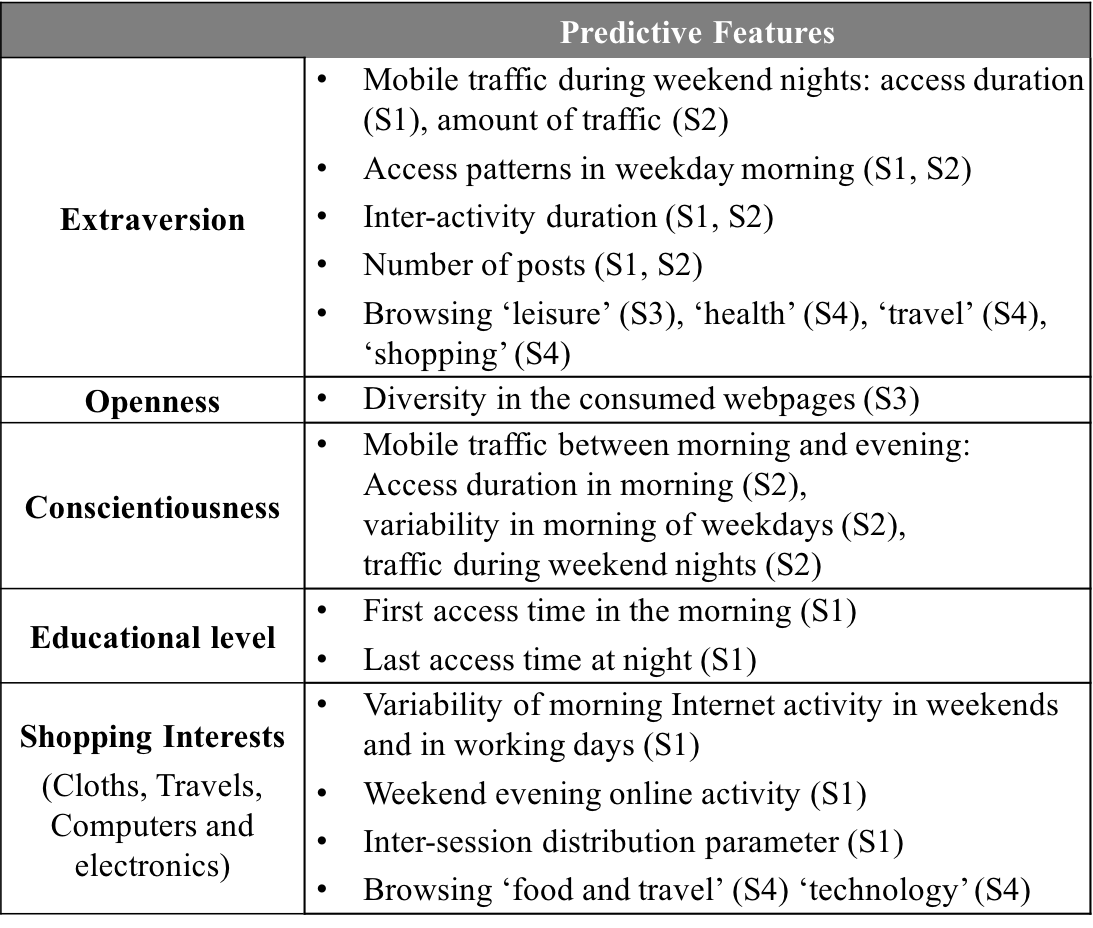}
  \caption{Overview of predictive features}
  \label{tbl:seven}
\end{table}
Next we provide a description of the most predictive features and interpret the associations between features and the target variables. Note that we are not able to provide explanations about the meaning of features computed from S3 data as they are produced through LSI, which makes the interpretation difficult. Therefore, we discuss the predictive power of S1, S2 and S4 features. 

\subsubsection{Personality Traits}
\paragraph{Extraversion} Extraverted individuals are defined as sociable, fun-loving and affectionate \cite{goldberg2006international}. The models built with features from all data types are able to classify individuals into \textit{high/low} extroversion with accuracies ranging between 66\% to 72\%. This result corroborates previous work where extraversion has also been inferred with high accuracy \cite{de2013predicting, staiano2014money}. In S1 and S2 data, the most predictive features are related to mobile activity during the night in the weekends (access duration in S1 data and amount of traffic in S2 data), the variability in access patterns in the morning during working days (S1 and S2 data), the distribution parameters of inter-activity durations (S1 and S2 data), and the number of "posts" (accessible only from S2 data). 

Several studies have associated extraversion and overall intensity of mobile phone usage (self-reported) as a means of stimulation \cite{carrascosa2015always, ramirez2010relationship}, which may be directly or indirectly reflected in our features. In addition, extroverts exhibit an increased usage of online leisure services \cite{amichai2008personality}, which may impact the weights of related website categories in S3 data. In terms of S4 data, topics related to health, travel and shopping seem to be the most predictive to classify extraversion; interestingly, the literature has reported an association between travel and shopping and extraversion \cite{hoxter1988tourist, wang2008passion}.

\paragraph{Conscientiousness} Highly conscientious individuals are characterized as efficient and organized \cite{goldberg2006international}. This trait was recognized the best with features from S2 data. The most predictive features capture the duration of morning activity, the amount of traffic during weekends in night hours, and the variability of morning activity during working days.  

The tendency of highly conscientious people to wake up earlier and to have specific diurnal preferences was reported two decades ago in the literature \cite{jiang2013calling} and has been consistently confirmed in recent years \cite{randler2008morningness} associating the characteristic named 'morningness' to this trait. In the technological literature, conscientiousness has been found to be negatively correlated with total Internet usage \cite{landers2006investigation}. Such characteristics of conscientious individuals might be captured in our experiments through the features that quantify morning vs. evening dynamics of mobile traffic, as well as the total traffic consumption. 

The fact that conscientiousness was not accurately classified with features from S1 data but with features from S2 data might suggest that this trait is strongly related to how people actually use the phone as opposed to general patterns of traffic. Recall that the main difference between these two data types is filtering out non-user initiated traffic. 

\paragraph{Openness} Openness to experience (often shortened as 'openness') depicts individuals who are creative, intellectual and insightful \cite{goldberg2006international}. Given this definition, the high classification accuracy of this personality trait from the features in S3 data (website category tags) may not be unexpected. As mentioned, due to the difficulty of reverse engineering the LSI-based features to website categories, we are unable to discuss the meaning of the most predictive features for S3 data (we had 1325 dimensions - categories - in this data when compared to 25 topics in S4 data). 

Openness has been found in the literature to be a predictor for a set of web services \cite{tsao2013big}. These correlations between openness and the types of websites support the accuracy of the models built with features from S3 data, which was higher than that of the models built with features from S4 data despite the fact that the latter relies on a more detailed analysis of the content of the web pages. It could be the case that openness is better captured by quantifying the diversity of the consumed webpages overall than performing in-depth topical analysis of the pages.  

\paragraph{Agreeableness and Neuroticism} Agreeableness reflects individual characteristics that are perceived as kind, sympathetic, cooperative, warm and considerate \cite{goldberg2006international}. Neuroticism is a trait that describes anxious, insecure, and self-pitying individuals \cite{goldberg2006international}. None of the four models was able to accurately infer agreeableness and neuroticism. We are careful in drawing an interpretation from this fact, since the reason might be in the limitation of our features and models, or in the fact that these traits are inherently difficult to capture with mobile online data. 

\subsubsection{Boredom Proneness}
Boredom Proneness was not accurately classified in our experiments. Recent work suggests that finer-grained information of smartphone usage might be necessary to infer this trait (e.g. screen-on events) beyond mobile HTTP(S) traffic \cite{matic2015boredom}.

\subsubsection{Demographics}
Automatic modelling of demographic information of unlabeled users has been also identified as an important user modelling task \cite{koenigstein2011yahoo}. Despite our expectations that the semantic information of browsed web pages would be predictive for this task, only the models built with features from S1 data were able to accurately classify individuals with low/high educational levels with 69\% balanced accuracy. The model relied mostly on features that quantified online activity in the first time periods in the morning and the last time periods in the night. We leave a more thorough investigation of these target variables and interpretation of results to future work. 

\subsubsection{Shopping Interests}
Inferring product interests with a small group of participants has clear limitations, especially considering the population scale that the major players of online behavioral advertising deal with. Our goal was not to target maximum accuracy but to explore the possibility of using different types of data for this purpose. Moreover, we have not found previous work aimed at inferring shopping interests from S1, S2 data that has neither information about websites nor content. Another research question that we were interested in exploring was whether mobile browsing logs (captured in features from S3 and S4 data) would have predictive power for this task. 

The most surprising finding was that two of the shopping categories --'clothes' and 'travel'-- were inferred from features from S1 data with 72\% and 64\% of balanced accuracy, respectively. The most indicative features for these two variables were related to the variability of morning Internet activity both in weekends and in working days, the weekend evening activity (only for 'travel') and the inter-session distribution parameters. Given the competitive performance in inferring extraversion from this data type, a possible explanation might be attributable to the link between extraversion and travel/shopping preferences observed in the literature \cite{huang2010relationship}.

Features from S3 and S4 data exhibited intuitive relations with the target variable, such as the relation of the topic 'food and travel' with the target variable 'travel' and the topic 'technology' with the target variable 'computer and electronics'.

\section{Implications}
\subsection{Possibility of Transparency and Negotiation}
As briefly discussed in the Introduction, we do not aim to draw one-sided interpretations from the results, i.e., either as  warnings about unexpected profiling threats or as solutions to more privacy-preserving profiling methods. We rather believe that the results are implying the possibility of transparency and negotiation between users and service providers. The availability of restricted views suggests a spectrum of compromises, rather than a binary decision where the users grant access to all of their data or to none of it. Previous work suggests \cite{leon2013matters} that explaining the scope of use and offering control to users can affect the willingness to share personal data, and our study provides concrete options for compromise. For example, an online ad exchange which can implement scenario 4 could transparently communicate scenario 3 (i.e., using up to domain names instead of full URLs) as an alternative. The users are then able to consider the trade-off in the relevance of advertisement and the degree of exposure of their browsing behavior. In addition, by choosing scenario 3 explicitly, users can have more trust that their full URL would not be exposed to the service provider. 

On the other hand, previous works showed a number of limitations of having transparency and providing end-users with privacy controls. Nissenbaum's work \cite{nissenbaum2011contextual} elaborates on the paradox of providing too much information to end-users, which discourages them to read and understand it (also make them avoid taking control of their privacy \cite{compano2010policy}), and that simplifying the complexity will inevitably leave out necessary details. Acquisti et al. \cite{acquisti2005privacy} also report various decision biases that often lead to short-term benefits while sacrificing long-term privacy. These works imply that privacy problems would not be addressed with simple transparency policies. We believe that the works further motivate the research on various related topics such as deliberate implementations of transparency goals, design of privacy regulations, development of decision support systems for end-users.

In addition, there is an opportunity for similar research in a variety of application domains, from existing online services and mobile apps to newly emerging ubiquitous systems that deal with highly personal data. For each application area, similar to our work, it will be important to identify the constraints in the application environment and analyze of the impact on the profiling capability. 

\subsection{It is not just about the Data}
With an increased awareness of privacy and the value of personal data, it is becoming important for Internet service providers to communicate to users about the use of personal data. The communication is often focused on explaining the types of collected data. The purpose of the collection is typically described in abstract terms. However, we have shown that it is possible to infer personal traits even from the data that are considered less sensitive (e.g. S1 data) when a proper technique is applied to the data. From the perspective of this work that puts emphasis on transparency, we believe that the communication should not be limited to the collected data but also convey the types of inferences that the service provider will make from the collected data and the purpose for making such inferences.

\subsection{The Power of Inferred User Profiles}
Though the inference of personal traits enables service providers to improve the personalization of their services (e.g., personality-aware friend recommendation \cite{bian2011online}), it may also open the door to manipulations of the users' behavior and decision making. For example, personality and boredom proneness are predictors of several behaviors, including impulsive online buying and gambling. This information may be misused by service providers to exploit human weaknesses to their advantage. Another example is Crystal, a startup that uses automatically inferred personality of email recipients to adapt the content and vocabulary of the answers
\footnote{Refer to https://www.theguardian.com/media/2015/may/19/crystal-knows-best-or-too-much-the-disconcerting-new-email-advice-service}
. These examples have already raised privacy and ethical concerns highlighted by the press. Our study suggests the need for clear codes of conduct by service providers to ensure an ethical treatment of the collected user data and associated inferences.

\subsection{The Importance of Dynamics in HTTP(S) Traffic}
In terms of understanding users through HTTP(S) traffic, significant research efforts have been devoted to inferring traits and preferences from the consumed content or visited destination. In our work, we observed that the time and frequency of HTTP(S) accesses also carry important information about users, such as the timestamp of HTTP(S) accesses (S1 data) enables the inference of certain personality traits (extraversion and conscientiousness), education level, and some shopping interests (clothing). This finding is timely as a large number of diverse services and devices have access such simple data. The results suggests that the potential of the data in terms of revealing personal information should be recognized by both collecting entities and users, and that it is important to clearly communicate the uses of the data.

\subsection{User Modeling in an HTTPS World}
The adoption of HTTPS by online services has been steadily increasing for the past years. Though the number varies depending on the measurement approach, it is commonly observed that a significant portion of the traffic is carried through HTTPS and the portion is growing \cite{naylor2014cost}. While one of the purposes of HTTPS is to protect the user's privacy, our results provide an interesting perspective about how much protection is achieved. According to our classification of the profiling constraints, HTTPS can be interpreted as a protocol to prevent S4 data, as it limits the view of the full URL and the consumed content. However, our results suggest that certain personality traits, education level and shopping interests can be inferred from S1-S3 data, which is visible for all HTTP(S) traffic. This should be considered in the evaluation of HTTPS, and HTTPS itself should not be understood as a full privacy solution. Again, the entities which have access to S3 data (e.g., ISP, middle boxes, WiFi access points) should recognize the sensitivity of the data and reflect on transparent and responsible uses. 
\section{Conclusions and Future Work}
In this paper, we have investigated the modeling capabilities and limitations of applying four different constraints to mobile HTTP(S) data. These constraints might arise due to a variety of reasons, including a lack of visibility of richer datasets by certain entities, and technical and/or legal limitations. We have presented the results of constrained user modeling on HTTP(S) data collected by means of an in-the-wild user study with 61 volunteers for at least 30 days. Our work shows that meaningful user traits and interests can be inferred from constrained data. We believe that our work contributes to the understanding of the trade-offs regarding user modeling, personalization and access to HTTP(S) data, particularly but not limited to a mobile context.

A direct extension of the work is to collect the data directly from mobile devices and conduct a similar study. Different types of data can be collected through the devices, for example, sensor readings, GPS positions, and usage of apps. Possible constraints can be defined in terms of accessing such data, and the impact of the constraints on user modeling can be analyzed. Recent literature suggests the importance of app usage \cite{seneviratne2014predicting}. Although we made an effort to identify app usage information from the HTTP(S) traffic and develop related features, we observed that the HTTP(S) traces do not accurately capture app usages; the domain name and the user-agent fields of the HTTP(S) logs were not enough to identify the app in many cases, and interactions with an app did not trigger an HTTP(S) connection often. Thus, we leave the exploration of the space for future work. 

In future work, it would be interesting to have access to all mobile Internet access data (including Wi-Fi) to carry out a comparative analysis. In addition, we would like to extend the data collection to PC devices at homes and conduct a similar comparative study. 

There are a number of user profile variables that our models could not infer, such as boredom proneness or gender. We plan to investigate this further and possibly design new features that might be more relevant to model these variables. Testing different  machine learning algorithms and conducting a concrete comparison would help understanding the profiling capability of the profiling scenarios. It can extend the findings of the associations between the target variables and the behavioral data. Moreover, observing consistent results across different algorithms would confirm the profiling capability of the scenarios.

Finally, it is important to explore new paradigms for transparency that would inform users not only about which data is being collected, but also how it is being analyzed and for which purposes. More in-depth studies on new negotiation models that can help both service providers and users are also necessary. 

\section{Acknowledgement}
The research leading to these results has received funding partially from the European Union's Horizon 2020 research and innovation programme under grant agreements No 653449 (project TYPES). The paper reflects only the authors' view and the Agency and the Commission are not responsible for any use that may be made of the information it contains.

\bibliographystyle{ACM-Reference-Format}
\bibliography{v1_references}

\end{document}